\documentclass[%
prd,
reprint,
floatfix,
aps,
longbibliography,
superscriptaddress,
nofootinbib]{revtex4-2}

\usepackage{amsmath,graphicx,amssymb,xcolor,booktabs,braket,multirow,cancel,enumerate,enumitem}
\usepackage[colorlinks=true, allcolors=blue]{hyperref}
\usepackage[normalem]{ulem}
\usepackage{soul}

\setstcolor{red}

\definecolor{Mahogany}{rgb}{0.75, 0.25, 0.0} 

\begin{document}

\title{
Quest for a phenomenologically consistent low cutoff theory 
}

\author{Sudhakantha Girmohanta}
\email{sgirmohanta@sjtu.edu.cn}
\affiliation{Tsung-Dao Lee Institute, Shanghai Jiao Tong University, Shanghai 201210, China}
\affiliation{School of Physics and Astronomy, Shanghai Jiao Tong University, Shanghai 201210, China}

\author{Yu-Cheng Qiu}
\email{ethan.qiu@cityu.edu.hk}
\affiliation{Tsung-Dao Lee Institute, Shanghai Jiao Tong University, Shanghai 201210, China}
\affiliation{Department of Physics, City University of Hong Kong, Kowloon, Hong Kong SAR, China}

\date{\today}

\begin{abstract}
The Randall-Sundrum model with the Higgs localized on the IR brane solves the gauge hierarchy problem. However, the associated low cutoff ($\Lambda \sim 10$ TeV) generically leads to unacceptably rapid nucleon decay and excessively large Majorana neutrino masses. Achieving consistency while simultaneously explaining the Yukawa hierarchy requires either a horizontal symmetry or a discrete gauged symmetry. We demonstrate that eliminating all dangerous operators within a horizontal symmetry framework must come with large and unattractive charge assignments, if possible at all. Hence, we consider an exact discrete gauged $\mathbf{Z}_N$ symmetry, with fermion mass hierarchies generated via wave function overlap. We employ this to reproduce the current Cabibbo-Kobayashi-Maskawa and Pontecorvo-Maki-Nakagawa-Sakata structures. Assuming universal five-dimensional Yukawa couplings, generation-blind profile for right-handed neutrinos and flat profile for the third generation SM doublets, it predicts Dirac neutrinos with a total mass $\sim 66$ meV. Since the $\mathbf{Z}_N$ charges must be generation blind, flavor observables serve as key probes.
\end{abstract}

\maketitle

\section{Introduction}

The naturalness problem associated with the Higgs mass can be addressed in a low cutoff framework, concretely realized via the Randall-Sundrum (RS) model~\cite{Randall:1999ee}. The RS model features one compactified extra dimension and warped bulk geometry, bounded by two $3$-branes (UV and IR brane) that reside on the fixed points of the extra dimension. The Standard Model (SM) Higgs is localized on the IR brane. The five-dimensional (5D) fundamental scale is close to the Planck scale, while all physical mass scales on the IR brane are redshifted by the warped geometry. Thus, the gauge hierarchy problem can be resolved by a stabilized extra dimension of suitable size~\cite{Goldberger:1999uk}. However, it simultaneously generates a low cutoff for higher-dimensional operators in the IR, denoted as $\Lambda$.\footnote{
For stabilization of the extra dimension, including the backreaction effects, see Refs.~\cite{Csaki:2000zn, Bellazzini:2013fga}. The radion can be somewhat lighter than $\Lambda$, and can be the first signature in a collider~\cite{Csaki:2007ns, Girmohanta:2023tdr}.}

In the Standard Model effective field theory approach, one should consider all possible Lorentz invariant operators that are allowed by the SM gauge symmetries $SU(3)_{\rm c} \times SU(2)_{\rm L} \times U(1)_{\rm Y}$ and suppressed by appropriate powers of $\Lambda$. This is problematic from the standpoint that a low quantum gravity scale violates all global symmetries, including the total baryon and lepton number. For example, the operator ${\cal Q} {\cal Q} {\cal Q} {\cal L}/\Lambda^2$ induces proton and (bound) neutron decay, where ${\cal Q}$ and ${\cal L}$ denote generic SM quark and lepton fields, respectively. Constraints on nucleon decay require $\Lambda \gtrsim 10^{16}$~GeV~\cite{Beneito:2023xbk}, which is hierarchically larger than the IR scale. Furthermore, the Weinberg operator $\ell H \ell H/\Lambda$~\cite{Weinberg:1979sa} gives unacceptably large neutrino masses after electroweak symmetry breaking, where $\ell$ denotes the $SU(2)_{\rm L}$ doublet lepton, and $H$ is the SM Higgs. Therefore, one needs additional structure for achieving consistency with observational constraints and to explain the observed Yukawa hierarchy.

One can use a generation-dependent horizontal symmetry~\cite{Froggatt:1978nt,Leurer:1992wg,Leurer:1993gy} of Froggatt-Nielsen (FN) type, whose spontaneous symmetry breaking (SSB) modulates the effective Yukawa couplings. The Yukawa hierarchy can be explained by the proper choice of FN charges for the SM fields~\cite{Qiu:2023igq}, while one can hope to suppress all dangerous nucleon decay operators concurrently.  However, we show in Appendix~\ref{appendix:FN} that very large and unattractive charge assignments are required to satisfy the proton decay bounds from only the dimension-six (dim-6) operators, while reproducing the correct Yukawa pattern. It appears challenging to simultaneously suppress all dangerous higher-dimensional baryon-number-violating operators and address the Yukawa hierarchy within the RS-FN framework with a given low cutoff $\Lambda$.

Another viable approach to suppress dangerous operators is to let SM fermions propagate in the bulk~\cite{Huber:2000ie, Gherghetta:2000qt, Chang:2005ya, Perez:2008ee, Agashe:2008fe, Iyer:2012db,Heidenreich:2014jpa}. Their zero-mode profiles in the extra dimension are controlled by the bulk mass parameters, and their overlap with the Higgs on the IR brane could generate the observed Yukawa hierarchy. This also raises the effective cutoff for higher-dimensional operators via the fermion wave function correction. This corresponds to renormalization suppression of the corresponding Wilson coefficients by large anomalous dimensions in the dual 4D conformal field theory~\cite{Maldacena:1997re}. However, sufficient suppression of nucleon decay operators in this setup typically oversuppresses Yukawa couplings, preventing a simultaneous explanation of the hierarchy~\cite{Gherghetta:2000qt, Huber:2000ie}.

In this work, we consider a unified solution to the above-mentioned problems in an RS-type low cutoff theory, which uses an exact discrete gauge $\mathbf Z_N$ symmetry~\cite{Ibanez:1991hv, Babu:2003qh, Davoudiasl:2005ks, Nakajima:2007uk} and bulk fermion profile. Imposing the exact $\mathbf Z_N$ with appropriate charges eliminates dangerous nucleon decay operators while allowing the Yukawa couplings. The Yukawa texture is generated via modulation from the SM fermion bulk profile, and we fit the observed quark and lepton mixing patterns. It predicts the Dirac neutrino with total mass $\sum_i m_i \simeq 66$~meV under the assumptions of the universal 5D Yukawa coupling, generation-blind profile for right-handed neutrinos, and flat profile for the third generation SM doublets.
Higher-order operators that induce flavor-violating effects are allowed by the generation-blind $\mathbf Z_N$, and we confront the theory with current constraints. Consistency with experimental constraints on lepton flavor-changing neutral current processes and electron electric dipole moment (EDM) with $\Lambda \sim \mathcal O( 10) $~TeV in this framework indicates additional flavor and $CP$ structures. We outline possible solutions. Since the discrete $\mathbf Z_N$ is exact, there is no associated domain wall problem in cosmology.


\section{Exact discrete \texorpdfstring{$\mathbf Z_N$}{ZN}}

We consider a low cutoff RS model with $\Lambda\sim \mathcal{O}(10)$~TeV with an exact discrete gauge $\mathbf Z_N$. 
With this choice of $\Lambda$, the mass of the lightest Kaluza-Klein (KK) gluon $\simeq (3/4)\pi \Lambda$ is compatible with the stringent bounds from flavor-changing neutral currents, including the constraints from $\epsilon_K$ measurements~\cite{Csaki:2008zd}, KK gauge boson mediated $\mu \to e \gamma$~\cite{Iyer:2012db}, and electroweak precision observables~\cite{Iyer:2015ywa}. For an exact $\mathbf Z_N$, to allow Yukawa couplings and mixing, one has to assign generation-blind charges
\begin{equation}
    n_Q^i = - n_{\bar u}^j = - n_{\bar d}^k = m\;,\quad n_\ell^i = - n_{\bar e}^j = n\;,
    \label{eq:mn}
\end{equation}
where $m,n<N$ are integers. $Q$ and $\ell$ are the $SU(2)_{\rm L}$ doublet and $\{\bar u, \bar d, \bar e\}$ are left-handed $SU(2)_{\rm L}$ singlets. The exact gauge $\mathbf Z_N$ has to be anomaly-free, i.e., instantons of the low-energy theory should not violate $\mathbf Z_N$. This is sufficient, as the explicit UV completion of such $\mathbf Z_N$ is not considered here~\cite{Banks:1991xj}.  Note that Eq.~\eqref{eq:mn} satisfies the anomaly-free condition of $\mathbf Z_N \times [SU(3)_{\rm c}]^2$ trivially, while the condition from $\mathbf Z_N\times [SU(2)_{\rm L}]^2$ implies
\begin{equation}
    \frac{3}{2}\left(3 m+n\right) \mod N = 0\;.
    \label{Eq:mnModN}
\end{equation}

Imposing the exact discrete $\mathbf Z_N$ on SM fermions can completely discard dangerous operators inducing nucleon decay. To discard dim-6 operators like $QQQ\ell$, one may assign $\mathbf Z_N$ charges only to leptons. However, this will allow dim-$9$ operators of the form ${\mathcal Q}^3 {\mathcal L}^3$, consistent with the anomaly cancellation condition in Eq.~\eqref{Eq:mnModN}. So the quarks also have to be charged under $\mathbf Z_N$.\footnote{While wave function profiles somewhat raise the effective cutoff, using the bulk profiles required for the fermion mass, we find that this class of operators is not consistent with experimental bounds and should be eliminated.} To have neutrino mass, one has to introduce right-handed neutrinos, whose charge is fixed by the $\mathbf Z_N \times [{\rm graviton}]^2$ anomaly-free condition, namely $n_{\bar \nu}^k = -n$. If $2n \mod N = 0$, the Weinberg operator $\ell H \ell H$ is allowed, which is incompatible with simultaneously having the observed charged lepton mass hierarchy and small neutrino masses. To avoid this, we require $2n \mod N \neq 0$, which also forbids Majorana mass terms for right-handed neutrinos. Thus, the simplest possibility is to consider Dirac neutrinos~\footnote{See Ref.~\cite{Nakajima:2007uk} for Majorana neutrino masses via $U(1)_X$ breaking on the Planck brane. Reference~\cite{Davoudiasl:2005ks} considered spontaneous breaking of gauged $\mathbf Z_N$, although it faces cosmological issues with domain walls and extra relativistic species.}.

With $\Lambda \sim \mathcal O(10)$ TeV, all baryon-number-violating operators having dimension less than $12$ should be eliminated. The minimal anomaly-free $\mathbf Z_N$ symmetry satisfying the above criteria, as well as forbidding the Weinberg operator, is $\mathbf Z_9$. The set of charge assignments consistent with anomaly cancellation and operator suppression is given by $(m, n) = (1,3)$. The lowest-dimensional operator allowed by this symmetry is $\mathcal Q^3 \mathcal L^5$, which can mediate exotic decay modes, such as invisible (bound) neutron decay~\cite{Heeck:2019kgr, Girmohanta:2019fsx}. The resulting lifetime is consistent with current bounds provided the effective cutoff scale is greater than $1.5$~TeV~\cite{SNO:2022trz}, and JUNO is expected to improve this sensitivity to $2$ TeV or more~\cite{JUNO:2024pur}. These can also mediate $p \to \ell^+ +{\rm inv.}$ decay modes, where $\ell^+ = e^+$, $\mu^+$, and the current experimental limit is satisfied for a cutoff scale larger than $2$~TeV~\cite{Super-Kamiokande:2014pqx}.\footnote{Our charge assignments also forbid $n$–$\bar n$ oscillations. Unlike split-fermion models, one can not separate quarks and leptons in the extra dimension to suppress nucleon decay in the RS model~\cite{Nussinov:2001rb, Girmohanta:2020llh}.}

Even though an exact gauge $\mathbf Z_N$ can prevent nucleon decay, it cannot reproduce the Yukawa texture (hierarchy) and still allows unsuppressed $\Delta F=2$ flavor-changing processes via higher-dimensional operators, namely, 
\begin{equation}
    ( \overline Q \gamma^\mu Q )^2\;,\quad \bar d Q \overline Q d \;,\quad \bar u Q \overline Q u\;,
    \label{eq:meson_operators}
\end{equation}
where the generation and $SU(2)$ indices are suppressed.
Their cutoffs should be $\gtrsim 100$~TeV~\cite{Isidori:2010kg}, which are not consistent with $\Lambda \sim \mathcal O (10)$~TeV.


\section{Wave function correction}

The Yukawa hierarchy and suppression of $\Delta F=2$ processes can be addressed with SM fermion fields in the bulk having a generation-dependent wave function profile. By a proper choice of the bulk profiles, we can reproduce the Yukawa texture and suppress $\Delta F=2$ processes~\cite{Agashe:2004ay,Agashe:2004cp,Csaki:2009wc}. We present a quantitative study in the following.

We consider the Higgs field locates on the IR brane.
The effective 4D Yukawa couplings for the zero modes after integrating out the extra dimension receive corrections from the fermion bulk wavefunction (profile) (detailed derivations are in Appendix~\ref{appendix:bulk}), which is 
\begin{equation}
    y^{ij} = y^{ij}_{(5)} q(c^i) q(c^j) \;, 
    \label{eq:yukawa_coupling}
\end{equation}
where $y^{ij}_{(5)}\sim \mathcal{O}(1)$ are the 5D Yukawa couplings with an anarchic distribution, and the wave function correction is
\begin{equation}
    q(c) = \sqrt{\left(\frac{1}{4}-\frac{c}{2}\right) \left(1+ \coth \left[ \left(\frac{1}{2} -c \right) \sigma_0\right] \right)} \;.
    \label{eq:q_c}
\end{equation}
Here $c \sim \mathcal{O}(1)$ is the dimensionless fermion bulk mass parameter, and $\sigma_0$ describes the size of the extra dimension, which in turn determines the 4D effective field theory (EFT) cutoff
\begin{equation}
    \sigma_0 = \log \frac{M_*}{\Lambda}\;.
    \label{eq:sigma}
\end{equation}
Here $M_*$ is the fundamental 5D scale, which is close to the reduced Planck scale, $M_*\simeq 2.4\times 10^{18}$~GeV.

Since the wave function corrections that contribute to the effective 4D Yukawa couplings can be factorized~\eqref{eq:yukawa_coupling}, the factor $q(c^j)$ is equivalent to the FN correction factor (reviewed in Appendix~\ref{appendix:FN}). One can easily find a set of choices on $c^j$ that can reproduce the observed Yukawa texture by utilizing the existing FN solutions. We solve the equation $q(c^j) = \lambda^{n_j}$,
to obtain the proper bulk parameter $c^j$ that gives the Yukawa texture for particle $j$, where $\lambda$ and $n_j$ are the corresponding FN parameter and FN charge that reproduces the observations. Taking Ref.~\cite{Qiu:2023igq} as an example, where $\lambda \simeq 0.171$, the wave function factors for the quark sector should be
\begin{align}
q(c_Q^i) & \simeq (\lambda^3, \lambda^2,1) \simeq (0.00500, 0.0292, 1.00) \;, \nonumber \\
q(c_u^i) & \simeq (\lambda^4, \lambda^{1.5}, 1) \simeq  (0.000855, 0.0707, 1.00) \;, \nonumber \\
q(c_d^i) & \simeq \frac{m_b}{m_t}(\lambda , 1, 1) \simeq (0.00294, 0.0172, 0.0172) \;.
\label{eq:q_quark}
\end{align}
There is an overall factor $m_b/m_t$ for the down sector compared to the up sector. We let $d$ carry it. As indicated in Ref.~\cite{Qiu:2023igq}, Eq.~\eqref{eq:q_quark} reproduces observed quark current masses and Cabibbo-Kobayashi-Maskawa (CKM) matrix. 

In our setup, neutrinos are Dirac. So the 4D Yukawa couplings for the lepton sector are $ y_\ell \bar \ell H e$ and $y_\nu \bar \ell \tilde H \nu$, where $y_\ell^{ij} \sim q(c_\ell^i) q(c_{e}^j)$ and $y_\nu^{ij} \sim q(c_\ell^i) q(c_{\nu}^j)$.
We diagonalize the charged lepton sector by $\ell \to U_\ell \ell$ and $e \to W_\ell e$, where $U_\ell^\dagger y_\ell W_\ell = {\rm diag} (m_e, m_\mu, m_\tau)/v$.
Decompose $U_\nu^\dagger y_\nu W_\nu = {\rm diag}(m_1,m_2,m_3)/v$, where $v\simeq 175$~GeV is the Higgs vacuum expectation value, and $W_\nu$ can be absorbed by $\nu \to W_\nu \nu$.
Then, the Pontecorvo-Maki-Nakagawa-Sakata (PMNS) matrix is $U_{\rm PMNS} = U_\ell^\dagger U_\nu$.

The choices on $\{c_\ell,c_{ e},c_{\nu}\}$ determine matrix elements of $y_\ell$ and $y_\nu$ up to an $\mathcal O(1)$ constant. 
Since $U_{\rm PMNS}$ is formed by unitary matrices that mix the $\ell$, the $q(c_\ell^j)$ sets the order of mixing angles. Here we only consider normal order, since it is preferred in approximate rank-one mass matrices.
From observation, we deduce that $q(c_\ell^1) \simeq 0.207$ and $ q(c_\ell^2) \simeq 0.942$, setting $q(c_\ell^3) \simeq 1.00$ (details are in Appendix~\ref{appendix:pmns}).
Accordingly, wave function correction for $e^i$ can be obtained from $q(c_{e}^i) \simeq  m_e/m_t q(c_\ell^i)$.
All correction factors are determined by the bulk mass parameter through Eq.~\eqref{eq:q_c}, summarized in Table~\ref{tab:c_j}.

Here we consider the simplest generation-blind bulk profiles for the right-handed neutrinos, $c_{\nu}^i = c_{\nu}$, which is a free parameter.~\footnote{This degenerate choice of bulk mass can be protected by symmetry. Here we use this simple choice to show that this framework can be used to make predictions based on certain assumptions.}
The 5D Yukawa couplings $y_{(5)}^{ij}$ are randomly sampled with magnitudes drawn from a Gaussian distribution centered at $1$ with a $0.1$ standard deviation and phases uniformly distributed between $0$ and $2\pi$. This yields scattered predictions suitable for comparison with experiments.

\begin{table}
    \caption{Bulk mass parameter $c$ for all SM fermions except right-handed neutrinos. Here we take $\Lambda = 10$~TeV.}
    \begin{ruledtabular}
    \begin{tabular}{c c c c}
        Generation $i$ &  $1$ & $2$ & $3$\\
        \midrule
       $c_Q$  &  $0.629$  &  $0.566$  & $-0.500$ \\
       $c_{u}$  & $0.688$ & $0.529$ & $-0.500$ \\
       $c_{d}$  & $0.648$ &  $0.586$ & $0.586$ \\
       $c_\ell $ & $0.460$ & $-0.387$  & $-0.500$ \\
       $c_{e}$  &  $0.821$ & $0.697$ & $0.604$ \\
    \end{tabular}
    \end{ruledtabular}
    \label{tab:c_j}
\end{table}
%

\begin{figure}
    \centering
    \includegraphics[width=7.4cm]{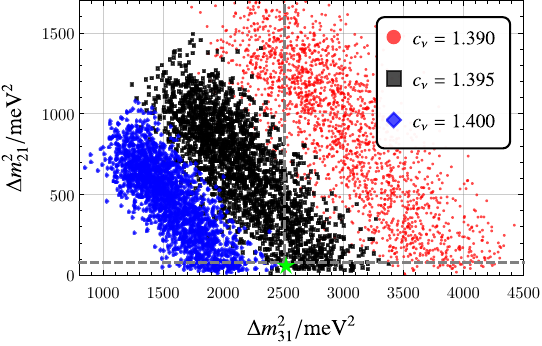}
    \includegraphics[width=7.4cm]{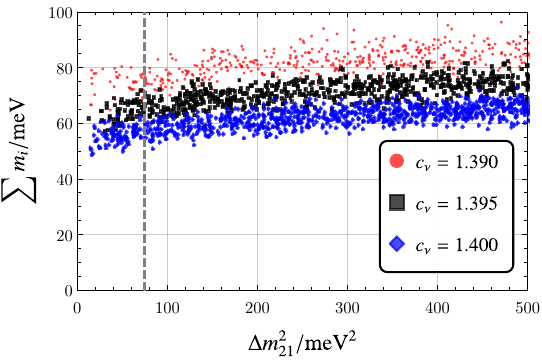}
    \caption{{Upper:} the prediction of neutrino mass-square differences under different choices of generation-blind $c_\nu$. Here, the normal ordering is considered. The horizontal and vertical dashed lines, whose intersection is labeled by a green star, are measured values adopted from Ref.~\cite{Esteban:2024eli}.
    {Lower:} the scattered predictions on the neutrino mass sum in the normal order with different choices of $c_\nu$. The vertical dashed line is the measured $\Delta m_{21}^2 =74.9^{+1.9}_{-1.9}$~meV$^2$~\cite{Esteban:2024eli}.
    }
    \label{fig:oscillation}
\end{figure}

As indicated in the upper panel of Fig.~\ref{fig:oscillation}, the neutrino mass-square differences show a clear correlation between the two observables, and $c_\nu = 1.395$ can provide a good fit to the measurements. 
Meanwhile, the sum of neutrino mass is also predicted and is shown in the lower panel. With the above assumptions,
we predict
\begin{equation}
    \sum_i m_i = 65.6 \pm 3.1 ~{\rm meV}\;,
\end{equation}
where the uncertainty is from that of $\Delta m_{21}^2$ and 5D Yukawa couplings. The error of predicted neutrino mass sum is $\sim 5\%$, which is smaller than the initial sampling of the $y_{(5)}^{ij}$. This is due to the cancellation between fluctuations of each element during the matrix diagonalization. Since Yukawa interactions are renormalizable, the specific choice of $\Lambda$ does not affect the prediction of the neutrino mass.

Let us calculate the effective suppression for operators that induce $\Delta F=2$ processes~\eqref{eq:meson_operators}. The effective cutoff scale for four-Fermi operators $\psi_I^i\psi_J^j\psi_K^k\psi_L^l/(\Lambda_{IJKL}^{ijkl})^2$ can be expressed compactly as
\begin{equation}
    \Lambda_{IJKL}^{ijkl} = \Lambda \frac{q\left(\frac{c_I^i+c_J^j+c_K^k+c_L^l-3}{2}\right)}{\sqrt{q(c_I^i) q(c_J^j) q(c_K^k) q(c_L^l)}} \;,
    \label{eq:Lambda_ijkl}
\end{equation}
where the capital letters label fermion species and small letters stand for generations.
To compare with experiments, we perform unitary rotations on all operators between generations to a proper basis.
So
\begin{equation}
    \frac{1}{\left(\Lambda_{IJKL}^{\alpha \beta \gamma \rho}\right)^2} =  \sum_{i,j,k,l} U_{I}^{i \alpha} U_{J}^{j \beta} U_{K}^{k \gamma} U_{L}^{l \rho} \frac{g_{IJKL}^{ijkl}}{\left(\Lambda_{IJKL}^{ijkl}\right)^2}   \;,
    \label{eq:Lambda_abcd}
\end{equation}
where $g_{IJKL}^{ijkl}$ is the Wilson coefficient, and $U_J$ labels the unitary transformation of the fermion field $J$, determined by Yukawa matrices.
For the charge conjugate field $I$, the corresponding $U^\dagger_I$ is used. 

\begin{figure*}
    \centering
    \includegraphics[width=14cm]{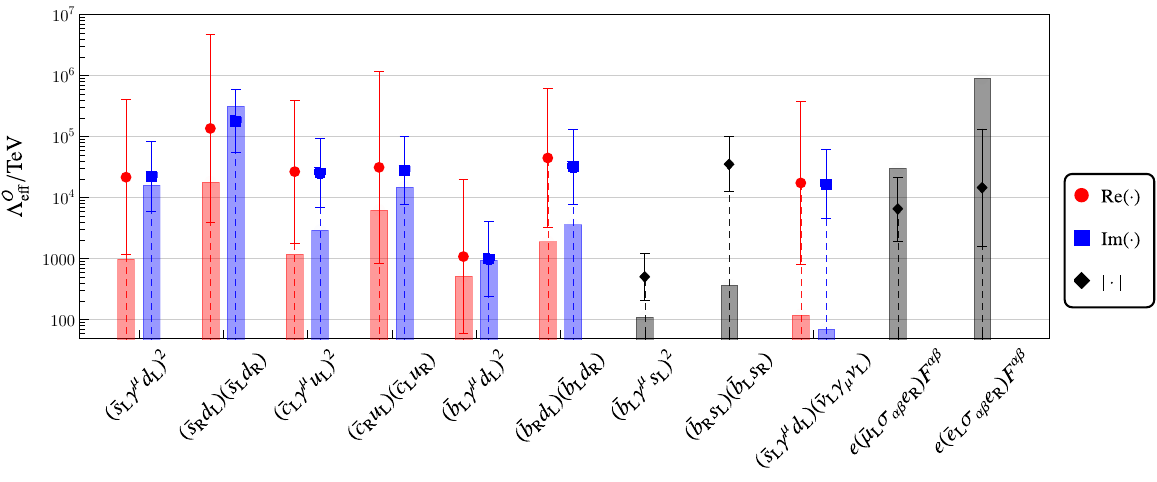}
    \caption{The prediction on the effective cutoffs of four-Fermi operators.
    The first eight operators from the left induce meson-antimeson oscillations~\eqref{eq:meson_operators}. The operator $(\bar s_{\rm L} \gamma^\mu d_{\rm L})(\bar \nu \gamma_\mu \nu)$ is responsible for the $K \to \pi \bar \nu \nu$. 
    The $\mu \to e \gamma$ is described by $e(\bar \mu_{\rm L} \sigma_{\alpha \beta} e_{\rm R}) F^{\alpha \beta}$. The last operator corresponds to the electron EDM.
    The solid bars are the existing bounds adopted from Refs.~\cite{Isidori:2010kg,Polonsky:2024pcc,Crivellin:2017rmk,Cesarotti:2018huy,Aebischer:2020dsw}. 
    The error bars of predictions are $3\sigma$ range in the log scale. 
    Here, all predictions are made by fixing $g_X^{\alpha\beta}=g_{IJKL}^{\alpha\beta \gamma \rho} =1$ and randomly sampling $\{g_X^{ij},g_{IJKL}^{ijkl}\}$ under the cutoff scale $\Lambda = 10$~TeV with bulk profiles in Table~\ref{tab:c_j}.
    }
    \label{fig:meson}
\end{figure*}

To make predictions, we first sample Wilson coefficients $g_{IJKL}^{ijkl}$ in the same way as $y_{(5)}^{ij}$.
Then, we generate unitary matrices for each fermion field using the bulk profile listed in Table~\ref{tab:c_j} and Eq.~\eqref{eq:q_c}.
Finally, with Eq.~\eqref{eq:Lambda_abcd},
we can predict the effective cutoff scales, which are presented in Fig.~\ref{fig:meson}.
All effective cutoffs for four-Fermi operators are consistent with current experimental bounds.

Note that for operator $(\bar s_{\rm R} d_{\rm L})(\bar s_{\rm L} d_{\rm R})$, it is marginally safe, which means that it can be probed or falsified in the future experiments.
Apart from meson-antimeson oscillation, we also present the bound from $K\to \pi \bar \nu \nu$, $\mu \to e \gamma$, and electron EDM.

\section{Dipole operators}

Here we consider the dim-6 dipole operators, among which the most constrained are operators inducing $\mu \to e \gamma$ and generating electron EDM. In the effective 4D theory, these operators are
\begin{equation}
    g_B \bar \ell H \sigma_{\mu\nu} e B^{\mu\nu}\;, \quad g_W \bar \ell H \sigma_{\mu\nu} \tau^a e W^{\mu\nu a} \;,
\end{equation}
where $g_{B,W}$ is the gauge coupling. To compare with experiments, one should perform unitary transformations according to Yukawa matrices and make a linear combination of the above operators according to electroweak SSB. We express $\mathcal O(1)$ 5D Wilson coefficients of above dipole operators as $y^{ij}_{\ell,(5)}+\Delta^{ij}$, which gives the effective cutoff as
\begin{align}
    \frac{\sqrt{2\sigma_0}}{(\Lambda_{\rm D}^{\alpha\beta})^2} & = \frac{1}{\Lambda^2} (U_\ell^\dagger)^{i \alpha}\left[  y_\ell^{ij} + \Delta^{ij}  q(c_\ell^i) q(c_e^j) \right] W_\ell^{j\beta} \nonumber \\
    & = \frac{m_e^{(\alpha) } \delta_{\alpha \beta}}{v \Lambda^2} + \frac{\Delta_{ij}}{\Lambda^2} (U_\ell^\dagger)^{i \alpha} q(c_\ell^i) q(c_e^j)  W_\ell^{j\beta}\;,
    \label{eq:Lambda_D}
\end{align}
where $\{U_\ell, W_\ell\}$ are unitary matrices used to diagonal charged lepton Yukawa matrix $y_\ell$.
The summation over $\{i,j\}$ is understood. 

There are two problems with a low cutoff theory concerning dipole operators, as shown in Fig.~\ref{fig:meson}.

\begin{figure}
    \centering
    \includegraphics[width=7.6cm]{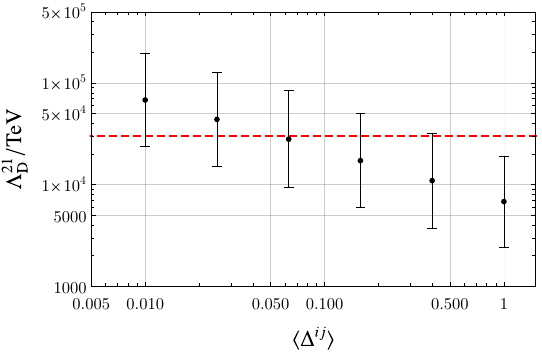}
    \caption{The prediction on effective cutoff scale on the tree level operator that induce $\mu \to e \gamma$ against assumptions on 5D Wilson coefficient $\Delta^{ij}$. The red dashed line is the current bound. We sample the $\Delta^{ij}$ with a Gaussian distribution centered at $\langle \Delta^{ij}\rangle$ whose standard deviation is $0.1 \times \langle \Delta^{ij}\rangle$ under the $\Lambda =10$~TeV. The error bars are $3\sigma$ range in the log scale.}
    \label{fig:Delta_ij}
\end{figure}

\begin{enumerate}
    \item The lepton flavor violation ($\alpha\neq \beta$) is determined by $\Delta^{ij}$.
    For $\mu \to e\gamma$, the vanishing branch ratio indicates that $\Lambda_{\rm D}^{21} \gtrsim 3 \times 10^4$~TeV~\cite{Crivellin:2017rmk}.
    Without additional flavor structure, one should generically have $\Delta^{ij}\sim \mathcal O (1)$.
    However, this cannot provide enough suppression for the $\mu \to e \gamma$ rate.
    Nevertheless, as shown in Fig.~\ref{fig:Delta_ij}, if we impose assumptions on $\Delta^{ij}$, one gets higher $\Lambda_{\rm D}^{21}$ for smaller $\langle \Delta^{ij}\rangle$.

    \item The electron EDM ($\alpha=\beta=1$) is determined by the imaginary part of $\Delta^{ij}$.
    Generally, ${\rm Im}(\Delta^{ij}) \sim \mathcal O(1)$ indicates that $\Lambda_{\rm D}^{(11)} \gtrsim 9\times 10^5$~TeV~\cite{Cesarotti:2018huy}. 
    This can be avoided by a spontaneous $CP$ violation model where a hidden sector is localized toward the UV brane and sources $CP$ violation, which is mediated to the CKM and PMNS matrices through SM light quarks and right-handed neutrinos, which are peaked toward the UV brane. As the dipole operator consists of the Higgs, which is localized on the IR brane, this would naturally suppress the electron EDM.
\end{enumerate}

Therefore, the above problems can be resolved either by raising the cutoff scale $\Lambda$ or through additional structure on lepton flavor and $CP$ of 5D Wilson coefficients.~\footnote{See for example Ref.~\cite{Aloni:2021wzk}.}

\section{Summary and Discussion}

In this work, we investigated a framework that explains the Higgs and Yukawa hierarchy with a stable nucleon. The naturalness of the Higgs scale is constructed by the warped extra dimension, which leads to a low cutoff theory in the effective 4D description. Our framework utilizes the exact discrete gauge $\mathbf Z_N$ symmetry to forbid dangerous nucleon decay operators, and predicts Dirac neutrinos. Here, the right-handed neutrino is necessary for the $\mathbf Z_N$ to be anomaly-free. Yukawa Hierarchy is provided by the fermion bulk profile. We provide a generation-blind choice of right-handed neutrino bulk profile that is consistent with current oscillation experiments and predicts the sum of neutrino mass to be $\sim 66$~meV assuming the universal 5D Wilson coefficients and flat profile of the third generation SM doublets.

Flavor observables can constrain this model further. Higher-dimensional operators that induce $\Delta F=2$ processes are allowed by exact $\mathbf Z_N$. We find that they are safe under the current bounds and may be probed in the future. However, dipole operators, which are also allowed by $\mathbf Z_N$, that induce $\mu\to e\gamma$ and electron EDM cannot escape current bounds. One may either raise the cutoff or impose additional flavor and $CP$ structure to make it consistent. We have outlined some of these possibilities.

Note that there exists a ``little hierarchy" between the cutoff $\Lambda\sim 10$~TeV and the measured Higgs scale $\sim 100$~GeV. The explanation of this $1\%$ tuning is beyond the scope of the current manuscript, since we are focusing on the phenomenological consistency of a low cutoff theory.

Different from broken $\mathbf Z_N$, this framework does not have the domain wall problem in cosmology. Electroweak baryogenesis or neutrinogenesis~\cite{Dick:1999je} can be easily synergized into our framework to produce matter-antimatter asymmetry without the need for heavy Majorana neutrinos. The exact $\mathbf Z_N$ can also provide absolute stability for a dark matter candidate. More detailed cosmological aspects of this framework, such as dark matter and dark energy, are pending future study.

\begin{acknowledgments}

We thank Yuichiro Nakai for helpful discussions.
The work of Y. -C. Q. is supported by the K. C. Wong Educational Foundation, GRF Grants No.~11302824 and No.~11310925, and CityUHK Grants No.~9610645 and No.~7020130. 

\end{acknowledgments}

\appendix

\section{FN solution to the nucleon decay}
\label{appendix:FN}

The FN solution to the Yukawa hierarchy problem is one possibility to suppress proton decay operators in a low cutoff theory. In the FN framework, a spontaneously broken (global or gauge) symmetry, continuous or discrete, is introduced. Let us take $U(1)_{\rm FN}$ as an example. Each SM fermion is assigned a charge under $U(1)_{\rm FN}$, and one introduces a scalar $\Phi$ having $-1$ charge.
Therefore, only certain operators that respect the FN symmetry are allowed.

For simplicity, here we consider the first generation only.
After the SSB of $U(1)_{\rm FN}$ by the vacuum expectation value (VEV) $\langle \Phi \rangle$, the effective Yukawa operators are given by
\begin{equation}
    \lambda_u \overline Q \tilde H u\;,\quad \lambda_d  \overline Q H d\;,\quad \lambda_e \overline \ell  H e \;,
    \label{eq:FN_yukawa}
\end{equation}
where $\lambda_u = \lambda^{|n_Q + n_{\bar u}|}$, $\lambda_d = \lambda^{|n_Q + n_{\bar d}|}$, and $\lambda_e = \lambda^{|n_\ell + n_{\bar e}|}$, where $n_J$ is the FN charge of particle $J$. Here, the FN parameter $\lambda = \langle \Phi \rangle/\Lambda<1$. 
To match the observation, one should have $\lambda_{u,d,e} \sim m_{u,d,e} / v$, where $v\simeq 175$~GeV is the Higgs VEV, which indicates that
\begin{align}
    |n_Q + n_{\bar u}| & \sim \frac{\log m_u/v}{\log \lambda}\;, \nonumber \\
    |n_Q + n_{\bar d}| & \sim \frac{\log m_d/v}{\log \lambda}\;, \nonumber\\
    |n_\ell + n_{\bar e}| & \sim \frac{\log m_e/v}{\log \lambda} \;.
    \label{eq:yukawa_constraint}
\end{align}
The order-one Wilson coefficient is neglected.
The above constraints reduce the degrees of freedom down to $2$. We take the independent charges to be $n_Q$ and $n_\ell$.

\begin{figure}
    \includegraphics[width=7cm]{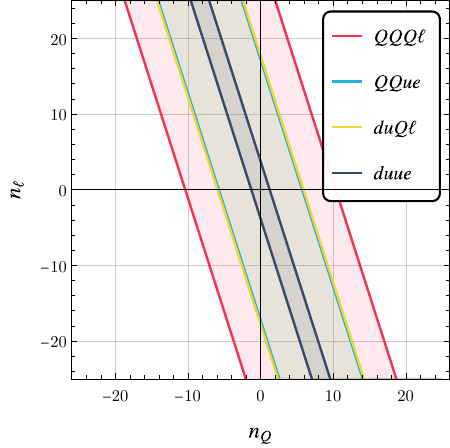}
    \caption{The parameter space for charges $n_Q$ and $n_\ell$ under the constraint from proton decay and Yukawa couplings~\eqref{eq:yukawa_constraint} for one generation only. Here we take the FN parameter $\lambda=0.17$~\cite{Qiu:2023igq} and the cutoff $\Lambda=10$~TeV. Shaded regions are excluded by proton decay, where $\Lambda_i$ in $n_{i,c}$~\eqref{eq:n_i,c} are taken from Ref.~\cite{Beneito:2023xbk}.}
    \label{fig:nq_nl}
\end{figure}

The effective dimension-six operators that lead to proton decay are
\begin{equation}
\lambda^{n_1} QQQ \ell \;,\quad \lambda^{n_2} QQue\;,\quad  \lambda^{n_3} d u Q \ell\;,\quad \lambda^{n_4} duue  \;,
\label{eq:dim-6_operators}
\end{equation}
where $n_j$ depends on the FN charges as follows:
\begin{align}
n_1 & = |3n_Q + n_\ell| \;, \nonumber\\
n_2 & = |2n_Q - n_{\bar u} - n_{\bar e}| \;, \nonumber \\
n_3 & = |-n_{\bar d} -n_{\bar u} + n_Q + n_\ell| \;,\nonumber\\
n_4 & = |-n_{\bar d} - 2 n_{\bar u} - n_{\bar e}| \;. \label{eq:dim-6_FN_charge}
\end{align}
%
The effective cutoffs are raised to $\Lambda \lambda^{-n_i/2}$, where $\Lambda$ is the cutoff of the EFT. To suppress nucleon decay to safety, one demands that $n_i > n_{i,\rm c}$, where
\begin{equation}
n_{i,\rm c} = \frac{2}{\log \lambda} \log \frac{\Lambda}{\Lambda_{i}}\;,
\label{eq:n_i,c}
\end{equation}
where $\Lambda_i$ is the experimental bound on the cutoff for the $i$th operator.
Using Eq.~\eqref{eq:yukawa_constraint}, which relates $\{n_{\bar u}, n_{\bar d}, n_{\bar e}\}$ to $\{n_Q, n_\ell\}$, the nucleon decay constraints translate to restrictions for charge assignments $n_Q$ and $n_\ell$. As shown in Fig.~\ref{fig:nq_nl}, one usually needs to take $n_{Q,\ell}\sim \mathcal{O}(20)$ to have a viable solution, which is unappealing.

The above analysis only considers the first generation. The SM consists of three generations, which means that Yukawa matrices are $3\times 3$ complex matrices. Since $Q$ is an $SU(2)$ doublet, one cannot simultaneously diagonalize the weak basis and the mass basis. Thus, SM has CKM mixing matrix, which is defined as $V_{\rm CKM} = U_u^\dagger U_d$, where $U_{u,d}$ are unitary matrices from the decomposition of quark Yukawa matrices,
\begin{equation}
    \lambda_u = U_u D_u W_u^\dagger\;,\quad \lambda_d = U_d D_d W_d^\dagger\;,
    \label{eq:diagonalization}
\end{equation}
where $D_{u,d}$ are diagonal real matrices, and unitary matrices $W_{u,d}$ can be absorbed by redefinition of $SU(2)$ singlets $u$ and $d$. Accordingly, the effective dimension-six operators that induce nucleon decay also involve different generations. For example, the effective Wilson coefficient of $Q^\alpha Q^\beta Q^\gamma \ell^\rho/\Lambda^2$ should be written as
\begin{equation}
    \lambda^{\alpha\beta\gamma\sigma}_{QQQ\ell} \sim \sum_{i,j,k,l} \lambda^{n_Q^i+n_Q^j+n_Q^k+n_\ell^l} U_u^{i\alpha} U_u^{j\beta} U_u^{k\gamma} U_\ell^{l\sigma}\;,
    \label{eq:lambda_abcd}
\end{equation}
where $n_{Q(\ell)}^i$ is the FN charge of $i$th generation of $Q(\ell)$ and $U_{u}$ is from Eq.~\eqref{eq:diagonalization}. 
$U_\ell$ is from the similar decomposition of $\lambda_e$. 
The Greek letters label the basis that can be used to compare with experiments. 
Meanwhile, $i$--$l$ labels the flavor basis where the FN charges are assigned. Here, $\mathcal{O}(1)$ Wilson coefficients are neglected for simplicity. 

The complete theory has in total $15$ FN charges to assign. The effective Weinberg operator $\ell H \ell H/\Lambda $ provides Majorana neutrino mass,
\begin{equation}
    m_\nu^{\alpha\beta} \sim U_\ell^{\alpha i}\lambda^{n_\ell^i + n_\ell^j} U_\ell^{j\beta } \frac{v^2}{\Lambda}\;,
\end{equation}
where the PMNS matrix is produced by diagonalizing $m_\nu$. The complete charge assignments that give the correct experimental measurement on the Yukawa texture can be obtained either through an analytical way~\cite{Qiu:2023igq} or a blanket search~\cite{Nishimura:2020nre,Cornella:2023zme,Nishimura:2024apb}. 
Using Eq.~\eqref{eq:lambda_abcd} to find a proper charge assignment that satisfies the proton decay bound is complicated, and one could expect that the resulting charges are large, as indicated in Fig.~\ref{fig:nq_nl}. Further, these are only the lowest-order dimension-six nucleon decay operators. To achieve consistency with the cutoff $\Lambda \sim 10$ TeV, one needs to suppress nucleon decay operators up to dimension 12. Therefore, we consider a different approach in the main text.


\section{The \texorpdfstring{$c_\ell$}{cl} and PMNS matrix }
\label{appendix:pmns}

The effective 4D Yukawa couplings for the lepton sector are
\begin{equation}
  y_\ell^{ij} \bar \ell^i H e^j \;,\quad y_\nu^{ij} \bar \ell^i \tilde H \nu^j\;,
  \label{eq:lepton_yukawa}
\end{equation}
where $y_\ell^{ij} = y_{\ell,(5)}^{ij} q(c_\ell^i) q(c_{e}^j)$ and $y_\nu^{ij} =y_{\nu,(5)}^{ij} q(c_\ell^i) q(c_{\nu}^j)$.
We decompose $U_\ell^\dagger y_\ell W_\ell = {\rm diag} (m_e, m_\mu, m_\tau)/v$ and $U_\nu^\dagger y_\nu W_\nu = {\rm diag}(m_1,m_2,m_3)/v$.
We diagonalize the charged lepton sector by $\ell^i \to U_\ell^{i \alpha} \ell^\alpha$ and $e^j \to W_\ell^{j\beta} e^\beta$, where $U_\ell^\dagger y_\ell W_\ell = {\rm diag} (m_e, m_\mu, m_\tau)/v$.
Then, the PMNS matrix is defined as $U_{\rm PMNS} = U_\ell^\dagger U_\nu$,
which connects the basis where the charged lepton sector is diagonalized and the neutrino mass eigenbasis $\ket{\nu_f} = U_{\rm PMNS} \ket{\nu_i}$. 

The standard parametrization~\cite{Giganti:2017fhf} of the PMNS matrix (without Majorana phase) is 
\begin{equation}
 U_{\rm PMNS} =  
\begin{pmatrix}
c_{12}c_{13}&s_{12}c_{13}&s_{13}e^{-i\delta}\\
\cdots & \cdots & s_{23}c_{13}\\
\cdots & \cdots & c_{23}c_{13}
\end{pmatrix}\;,\label{eq:PMNS}
\end{equation}
where $s_{ij} = \sin \theta_{ij}$ and $c_{ij} = \cos \theta_{ij}$. Meanwhile, in the limit $y_{(5)}^{ij}\to 1$, we have
\begin{align}
&U_\ell \sim U_\nu \to \label{eq:approximate_U} \\
& \begin{pmatrix}
-\frac{q_1 q_2}{\sqrt{(q_1^2+q_3^2)(q_1^2+q_2^2+q_3^2)}} & - \frac{q_3}{\sqrt{q_1^2+q_3^2}} & \frac{q_1}{\sqrt{q_1^2 +q_2^2 + q_3^2}} \\ 
\cdots & \cdots & \frac{q_2}{\sqrt{q_1^2 + q_2^2 + q_3^2}} \\
 \cdots & \cdots & \frac{q_3}{\sqrt{q_1^2 + q_2^2 + q_3^2}}
\end{pmatrix}\;,\nonumber
\end{align}
where $q_j = q(c_\ell^j)$. So one can estimate $q(c_\ell^j)$ by using the observed $\theta_{ij}$ and comparing Eqs.~\eqref{eq:PMNS} and~\eqref{eq:approximate_U}, which gives
\begin{align}
    \tan \theta_{23}  &\sim \frac{q(c_\ell^2)}{q(c_\ell^3)}  \;, \\
    \sin \theta_{13}  &\sim \frac{q(c_\ell^1)}{\sqrt{q(c_\ell^1)^2+q(c_\ell^2)^2+q(c_\ell^3)^2}} \;.
    \label{eq:estimate}
\end{align}
Taking global analysis from Ref.~\cite{Esteban:2024eli}, IceCube/DeepCore (IC24) with Super-Kamiokande atmospheric data, we obtain $q(c_\ell^1) \simeq 0.207$ and $ q(c_\ell^2) \simeq 0.942$, setting $q(c_\ell^3) \simeq 1.00$.

\begin{figure}
    \centering
    \includegraphics[width=8cm]{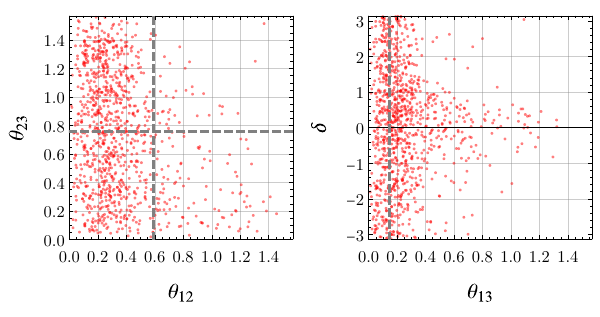}
    \caption{The scattered predictions on the PMNS mixing angles and $CP$ phase in the normal order and with $c_\nu=1.395$. The horizontal and vertical dashed lines are measured values adopted from Ref.~\cite{Esteban:2024eli}. 
    Note that the choice on $c_\nu$ does not affect the PMNS parameters here much, since they are mostly determined by $c_\ell$.}
    \label{fig:PMNS}
\end{figure}

As a cross-check, taking $c_\nu =1.395$, one can obtain the prediction of the PMNS matrix parameters, which is shown in Fig.~\ref{fig:PMNS}, and they are consistent with observations.

%


\section{Bulk fermion and effective cutoff}
\label{appendix:bulk}

The RS background describes a $\mathbf{R}_4 \otimes S^1/{\mathbf{Z}_2}$ slice of AdS$_5$, where the background metric is
\begin{equation}
    ds^2 = e^{-2 \sigma(y)} \eta_{\mu \nu} dx^\mu dx^\nu + d y^2 \equiv g_{MN} dx^M dx^N \ ,
    \label{Eq:ds2}
\end{equation}
where $\mu, \nu=0,\cdots,3$, while $M, N$ denote a generic 5D coordinate, $y$ stands for the extra-dimensional direction, $\eta_{\mu \nu} = {\rm diag}(-1,+1,+1,+1)$, $\sigma(y)=k |y|$, and $k$ represents the anti-de Sitter (AdS) curvature. The extra dimension is bounded by two 3-branes at the fixed points $y=0, r_0$, denoted as the UV and IR branes, respectively. We denote $\sigma(r_0)$ as $\sigma_0$. The extra dimension can be stabilized by the Goldberger-Wise mechanism~\cite{Goldberger:1999uk}. The four-dimensional effective Planck scale $M_{\rm Pl}$ is related to the fundamental five-dimensional Planck scale $M_*$ as 
\begin{equation}
    M_{\rm Pl}^2  = \frac{M_*^3}{k} \left( 1-e^{-2\sigma_0} \right) \;.
\end{equation}
Because of the nonfactorizable geometry, $M_* \sim M_{\rm Pl}$, while all the physical mass scales on the IR brane are redshifted by the warp factor $\Lambda = M_* e^{-\sigma_0}$. This can therefore address the gauge hierarchy problem.

A general bulk 5-dimensional Dirac fermion $\Psi(x^\mu,y)$ has the action
\begin{align}
    S_{\Psi} = -\int d^5x \sqrt{|g|} \  k \left[ \overline \Psi \gamma^N e_{N}^{M} \left( \partial_M + \omega_M \right) \Psi + m_\Psi \overline \Psi \Psi \right] \ ,
\end{align}
where $g$ denotes the determinant of the metric, the vielbein $e^M_N = {\rm diag}(e^\sigma, e^\sigma, e^\sigma, e^\sigma,1)$, $\gamma_M$ satisfies the 5D Dirac algebra in Minkowski space, the spin connection $\omega_M = (-\frac{k}{2}e^{-\sigma} \gamma_\mu \gamma^5,0)$, and $m_\Psi = c k$ denotes the bulk mass. The dimensionless parameter $c$ determines the 5D profile of the fermion. A factor of $\sqrt{k}$ has been extracted such that $\Psi$ carries mass dimension $3/2$.
Decomposing the Dirac spinor $\Psi = \Psi_{+}+ \Psi_{-}$, where $ \Psi_{\pm} = \mp \gamma^5 \Psi_{\pm}$, and performing the KK decomposition for each of the chiral components
\begin{equation}
    \Psi_{\pm} (x^\mu,y) = \sum_{p=0}^{\infty} \psi_{\pm}^{(p)}(x^\mu) f^{(p)}_{\pm}(y) \ ,
\end{equation}
and solving the equation of motion for the zero mode, one obtains that one of the chiral modes is always eliminated by the boundary condition, and 
\begin{equation}
    f_{\pm}^{(0)}(y) = d_{\pm}^{(0)} e^{(2\mp c) \sigma(y)} \ ,
    \label{Eq:fpm}
\end{equation}
where the normalization constant $d_{\pm}^{(0)}$ is obtained by canonically normalizing the zero-mode kinetic term. For the left-handed zero mode 
\begin{align}
    S_\Psi & \supset - \int d^4 x \left(d_+^{(0)}\right)^2 \mathcal I(c)  \bar\psi_+^{(0)} \gamma^\mu \partial_\mu \psi_+^{(0)} + \cdots  \;,
\end{align}
where
\begin{equation}
\mathcal I(c) = 2  \int_0^{\sigma_0} d\sigma e^{(1 - 2c ) \sigma } = \frac{2\left(  1-e^{(1-2c)\sigma_0}  \right)}{2c-1} \;,
\label{Eq:IcDef}
\end{equation}
and the prefactor $2$ is due to orbifolding. To canonically normalize the kinetic term, $d_+^{(0)} = 1/\sqrt{\mathcal I(c)}$. Notice, ${\mathcal I}(c)$ is always positive. For $c>1/2$ ($c<1/2$), the zero mode is localized toward the UV (IR) brane with respect to the 5D flat metric. Similar results hold for the right-handed mode when $c \to -c$.

SM Weyl fermions of a given chirality $\psi_{i+}$ ($\psi_{i-}$) are identified with the corresponding zero-mode of a 5D Dirac spinor $\Psi_{i}^{({\rm L})}$ ($\Psi_{i}^{({\rm R})}$) with appropriate choice of the boundary conditions. Here $i$ denotes generations. The SM Higgs field is localized on the IR brane,
\begin{align}
    \nonumber
    S_{H} & \supset - \int d^5 x \sqrt{|g|} \left(  g^{\mu \nu} \partial_\mu H^\dagger \partial_\nu H  + \cdots \right)  \delta(y - r_0) \\
    & = - \int d^4 x \left( e^{-2 \sigma_0} \eta^{\mu \nu} \partial_\mu H^\dagger \partial_\nu H + \cdots \right) \ ,
\end{align}
hence to canonically normalize its kinetic term one redefines $H \to e^{\sigma_0} H$. Therefore, the Yukawa interactions are of the form
\begin{align}
\nonumber
& \int  d^5x \sqrt{|g|} y_{ij}^{(5)}   \left[ \overline \Psi_i {}^{({\rm L})} \Psi_j^{({\rm R})} + {\rm H.c.}\right] H \delta(y-r_0) \\
& \to \int d^4x \  y_{ij}  \bar \psi_{i+}^{(0)} \psi_{j-}^{(0)} H + {\rm H.c.} + \cdots  \ .
\end{align}
Hence, the effective 4D Yukawa couplings are
\begin{equation}
y_{ij} = y_{ij}^{(5)} \frac{e^{(1- c_i - c_j) \sigma_0}}{\sqrt{\mathcal I (c_i) \mathcal I(c_j) }} = y_{ij}^{(5)} q(c_i) q(c_j) \;,
\end{equation}
where $c_i^{({\rm L})} = c_i$, $c_j^{({\rm R})} = -c_j$, and
\begin{align}
    q(c) & \equiv \frac{e^{\kappa(c)}}{\sqrt{{\mathcal I}(c)}} = \sqrt{\frac{\kappa(c)}{2\sigma_0} \left[1+ \coth \kappa (c) \right]} \ ,
    \label{Eq:defqc}
\end{align}
where $\kappa(c) = (1/2-c)\sigma_0$. 
As the $c$ coefficient for the right-handed field can always be redefined as $c_{\rm R} \to -c_{\rm R}$, we can use the profile for $\psi_{+}^{(0)}$ for a given operator without loss of generality.

The 4D effective dim-6 four-Fermi operators originate from the corresponding 5D operators as
\begin{align}
S_{\rm eff}^{(4)} & \supset \int d^5x \sqrt{|g|}  k \frac{ \Psi_i \Psi_j \Psi_k \Psi_l }{M_*^2} \\
& =\int d^4x  \int_0^{\sigma_0} d\sigma \frac{2 \  e^{(4-c_i-c_j-c_k- c_l )\sigma}}{\sqrt{\mathcal I (c_i) \mathcal I (c_j) \mathcal I (c_k) \mathcal I (c_l)}} \nonumber\\
&\qquad \times \frac{\psi_{i+}^{(0)} \psi_{j+}^{(0)} \psi_{k+}^{(0)} \psi_{l+}^{(0)}}{M_*^2} + \cdots \\
& =\int d^4x  \frac{\psi_{i+}^{(0)} \psi_{j+}^{(0)} \psi_{k+}^{(0)} \psi_{l+}^{(0)}}{\Lambda_{ijkl}^2} + \cdots \;,
\end{align}
where Eq.~\eqref{Eq:fpm} is used. Therefore, the effective 4D cutoff in the low-energy theory for the above four-Fermi interaction $\Lambda_{ijkl}$ is
\begin{equation}
\left(\frac{M_*}{\Lambda_{ijkl}} \right)^2 =  2 \int_0^{\sigma_0} d\sigma \frac{ e^{(4-c_i-c_j-c_k- c_l )\sigma} }{\sqrt{\mathcal I (c_i) \mathcal I (c_j) \mathcal I (c_k) \mathcal I (c_l)}} \;. 
\end{equation}
Using Eqs.~\eqref{Eq:IcDef} and~\eqref{Eq:defqc}, the above equation can be expressed as a combination of wave function correction factors as in Eq.~(9) in the main text. Moreover, this can be generalized for an effective operator involving $p$ fermions
\begin{align}
S_{\rm eff}^{(p)} & \supset \int d^5x \sqrt{|g|}  k \frac{ \Psi_1 \Psi_2 \cdots \Psi_p }{M_*^{(3p/2-4)}} \\
& =\int d^4x  \frac{\psi_{1+}^{(0)} \psi_{2+}^{(0)} \cdots \psi_{p+}^{(0)}}{\Lambda_{p}^{(3p/2-4)}} + \cdots \;,
\end{align}
and the effective 4D cutoff scale $\Lambda_p$ is evaluated to be
\begin{align}
    \Lambda_p = \Lambda \left[ \frac{q\left( \frac{ (5-2p) + \sum_{i=1}^{p} c_i }{2}\right)}{ \prod_{i=1}^{p}\sqrt{ q\left( c_i\right)}} \right]^{\frac{4}{3p-8}} \ ,
    \label{Eq:generalp}
\end{align}
where $\Lambda = M_* e^{-\sigma_0}$ is the IR cut-off scale. Eq.~(9) in the main text is a special case of Eq.~\eqref{Eq:generalp} for $p=4$.

Let us also consider the effective cutoff for dipole operators of the form
\begin{align}
    \nonumber
    S_{\rm eff}^{(D)} & \supset \int d^5x \sqrt{|g|} \frac{1}{M_*^2} \overline \Psi_i \sigma^{MN} \Psi_j F_{MN} H \delta(y-r_0) \\ 
    & \to \int d^4x \frac{1}{\Lambda_{D, ij}^2} \overline \psi_{i+}^{(0)} \sigma^{\mu \nu} \psi_{j-}^{(0)} F_{\mu \nu}^{(0)} H + \cdots \ .
\end{align}
Taking $H \to e^{\sigma_0} H$ rescaling, normalization of the zero-mode gauge boson $1/{\sqrt{2 \sigma_0}}$, and the vielbein involved in the $\sigma^{MN}$ definition into account, we obtain the 4D effective cut-off as
\begin{equation}
   \Lambda_{\rm D}^{ij} = \Lambda \left[ {\frac{q(c_i) q(c_j)}{\sqrt{2\sigma_0}}} \right]^{-1/2} \ .
\end{equation}
Finally, let us note the cutoff for a nonrenormalizable operator involving $p$ fermions and $s$ Higgs fields
\begin{align}
     S_{\rm eff}^{(ps)} & \supset \int d^5x \sqrt{|g|} \frac{\Psi_1 \Psi_2 \cdots \Psi_p H^s}{M_*^{(3p/2)+s-4}}  \delta(y-r_0) \ .
\end{align}
The effective cutoff is found to be
\begin{equation}
    \Lambda_{(ps)} = \Lambda \left[ \prod_{i=1}^{p} q(c_i) \right]^{\frac{2}{3p+2s-8}} \ .
\end{equation}

\bibliography{ref}

\end{document}